\documentclass[superscriptaddress,aps,twocolumn,longbibliography,nofootinbib]{revtex4-2}
\pdfoutput=1
\usepackage{graphicx, color, graphpap}      % Include figure files
\usepackage{amsmath}
\usepackage{enumitem}
\usepackage{amssymb}
\usepackage{amsthm}
\usepackage{pstricks}
\usepackage{float}
\usepackage[colorlinks=true, citecolor=blue, linkcolor=red]{hyperref}
\usepackage[T1]{fontenc}
\usepackage{bbm}
\usepackage{dsfont}
\usepackage[linesnumbered,ruled,vlined]{algorithm2e}
\SetKwInput{kwInit}{Init}
\usepackage{mathtools}

\usepackage{tikz}
\usepackage{graphicx}
\usetikzlibrary{positioning}
\usepackage{graphicx}
\usepackage{xcolor}
\usepackage[caption=false]{subfig}
\usepackage[ruled,vlined]{algorithm2e}
\usepackage{siunitx}
\usepackage{qcircuit}
\usepackage{pifont} % http://ctan.org/pkg/pifont


\renewcommand{\eqref}[1]{(\ref{#1})}
\newtheoremstyle{example}{\topsep}{\topsep}%
{}%         Body font
{}%         Indent amount (empty = no indent, \parindent = para indent)
{\bfseries}% Thm head font
{:}%        Punctuation after thm head
{   }%     Space after thm head (\newline = linebreak)
{\thmname{#1}\thmnumber{ #2}}%\thmnote{ #3}}%         Thm head spec
\theoremstyle{example}
\theoremstyle{definition}

\newtheorem*{theorem*}{Theorem}

\def\orcid#1{\kern -0.4em\href{https://orcid.org/#1}{\includegraphics[keepaspectratio,width=0.7em]{orcid_logo.pdf}}}

\usepackage{lipsum}

\long\def\ca#1\cb{} %Use for commenting out: \ca...\cb
\begin{document}
\title{Error mitigation increases the effective quantum volume of quantum computers}

\author{Ryan LaRose}
\thanks{Corresponding author: \href{mailto:ryan@unitary.fund}{ryan@unitary.fund}.}
\affiliation{Department of Computational Mathematics, Science, and Engineering, Michigan State University, East Lansing, MI 48823, USA}
\affiliation{Unitary Fund}

\author{Andrea Mari}
\affiliation{Unitary Fund}

\author{Vincent Russo}
\affiliation{Unitary Fund}

% \author{Sarah Kaiser}
% \affiliation{Unitary Fund}

\author{Dan Strano}
\affiliation{Unitary Fund}

\author{William J. Zeng}
\affiliation{Unitary Fund}
\affiliation{Goldman, Sachs \& Co, New York, NY}

\begin{abstract}
    Quantum volume is a single-number metric which, loosely speaking, reports the number of usable qubits on a quantum computer. While improvements to the underlying hardware are a direct means of increasing quantum volume, the metric is ``full-stack'' and has also been increased by improvements to software, notably compilers. We extend this latter direction by demonstrating that error mitigation, a type of indirect compilation, increases the effective quantum volume of several quantum computers. Importantly, this increase occurs while taking the same number of overall samples. We encourage the adoption of quantum volume as a benchmark for assessing the performance of error mitigation techniques. 
\end{abstract}

% =============================================================================
% =============================================================================
\maketitle
% =============================================================================
% =============================================================================

\textbf{Introduction} \ \ 
Quantum volume~\cite{qv} is a single-number metric which, loosely speaking, reports the number of usable qubits on a quantum computer\footnote{Some authors define quantum volume as the  effective Hilbert space dimension. In this paper we report the logarithm of this number which corresponds to the number of qubits.}. While improvements to the underlying hardware are a direct means of increasing quantum volume, the metric is ``full-stack'' and can be increased by an improvement to any component, e.g. software for compilation to produce an equivalent quantum circuit with fewer elementary operations~\cite{qv_compile}.

Given an $m$ qubit quantum circuit $C$, the \textit{heavy set} is $\mathcal{H}_C := \{ z \in \{0, 1\}^m : p(z) > p_\text{median} \}$ where $p(z) := \left | \langle z | C | 0 \rangle \right |^2$ is the probability of sampling bitstring $z$ and $p_\text{median}$ is the median probability over all bitstrings. A \textit{heavy bitstring} is one in the heavy set. Quantum volume is determined by counting the number of heavy bitstrings $n_h$ measured over $n_c$ random circuits, each sampled $n_s$ times. If the experiment is run with $m$ qubit circuits of depth $d = m$, $n_c \ge 100$, and
\begin{equation} \label{eqn:volume-condition}
    \hat{h}_d := {n_h} / {n_c n_s}  > 2 / 3 + 2 \sigma  
\end{equation}
where $\sigma$ is the standard deviation of the estimate, then the volume is at least $m$. The actual volume is the largest $m$ such that these conditions are true. The particular structure of these random circuits, which we refer to as \textit{quantum volume circuits}, is defined in~\cite{qv}.

\textbf{Method} \ \ 
Given a quantum volume circuit $C$, we define the projector on the heavy subspace 
\begin{equation}
    \Pi_{h, C} := \sum_{z \in \mathcal{H}_C} |z\rangle \langle z|
\end{equation}
so that the expected number of heavy bitstrings for this circuit is $n_{h, C} := n_s \langle 0 | C^\dagger \Pi_{h, C} C | 0 \rangle$. We use zero-noise extrapolation (ZNE)~\cite{zne_simon, zne} with $\Pi_{h, C}$ as the observable for each quantum volume circuit $C$ to estimate the noise-free value of $n_h := \sum_C n_{h, C}$. This amounts to evaluating $\langle \Pi_{h, C}^{(\lambda)} \rangle$ at several noise-scale factors $\lambda \ge 1$ then using these results to estimate $\langle \Pi_{h, C}^{(0)} \rangle$, i.e., the zero-noise limit of the heavy output probability. In practice, this means compiling the circuit $C$ to a set of circuits $\{ C_{\lambda_i} \}_{i = 1}^{k}$. For fairness with the unmitigated experiment, we use $n_s / k$ samples for each $C_{\lambda_i}$ so that the total number of samples drawn is equal in the mitigated and unmitigated experiments. The main difference to previous work improving quantum volume by compiling~\cite{qv_compile} is that we compile a single circuit to a set of circuits, following the pattern of many error mitigation methods (e.g.~\cite{zne}), in contrast with compilation that does rewrites on a single circuit following algebraic rules or optimized routing.

After executing each $C_{\lambda_i}$ to obtain scaled heavy output counts $n_{h, C}^{(i)}$, we use Richardson extrapolation~\cite{zne, dzne} to estimate the zero-noise result via
\begin{equation} \label{eqn:zne-result}
    n_{h, C}^{(0)} = \sum_{i = 1}^{k} \eta_i n_{h, C}^{(i)} 
\end{equation}
where coefficients are given by
\begin{equation}
    \eta_i := \prod_{j \neq i} \frac{\lambda_j}{\lambda_j - \lambda_i} .
\end{equation}
In practice, we use $\lambda_i \in \{1, 3, 5, 7, 9\}$ and scale circuits by locally folding two-qubit (CNOT) gates~\cite{dzne}. In other words, the scaled circuit for $\lambda_i = t$ has each CNOT replaced by $t$ CNOTs.

\begin{figure*}
    \centering
    \includegraphics[width=2\columnwidth]{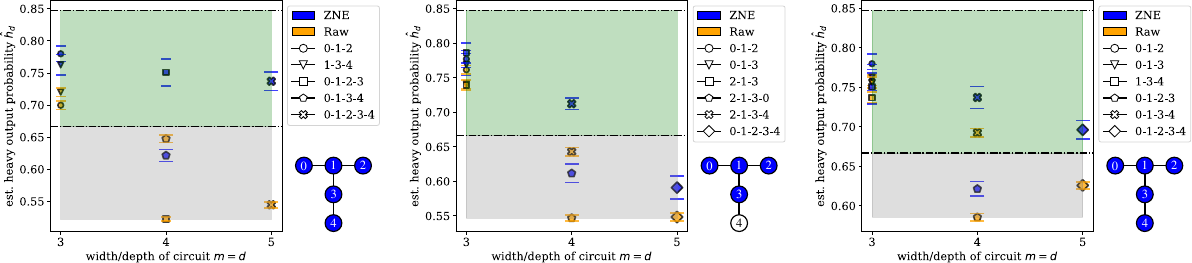}
    \caption{Results of unmitigated and mitigated quantum volume experiments on three five-qubit quantum computers (left-to-right: Belem, Lima, and Quito) using $n_c = 500$ circuits and $n_s = 10^4$ total samples. Each marker shows the estimated heavy output probability $\hat{h}_d$ on a different qubit configuration defined in the legend and error bars show $2\sigma$ intervals evaluated by bootstrapping. The connectivity of each device is shown below each legend. Dashed black lines show the $2/3$ threshold and noiseless asymptote $(1 + \ln 2) / 2$~\cite{qv}. For the mitigated experiments, $\lambda_i \in \{1, 3, 5, 7, 9\}$ and $n_s = 10^4 / 5$. Local unitary folding of two-qubit gates is used to compile the circuits (i.e., scale noise) and Richardson's method of extrapolation is used to infer the zero-noise result. The qubit subsets which achieved the largest quantum volume in the mitigated experiments are colored blue in each device diagram. As can be seen, on Belem error mitigation increases the effective quantum volume from three to five, on Lima error mitigation increases the effective quantum volume from three to four, and on Quito error mitigation increases the effective quantum volume from four to five.
}
    \label{fig:main}
\end{figure*}

\textbf{Results} \ \ 
Using this strategy, we perform unmitigated and mitigated quantum volume experiments on the Belem, Lima, and Quito devices available through IBM~\cite{ibmq} (see Appendix~\ref{app:device-specifications} for device specifications). The results, shown in Fig.~\ref{fig:main}, demonstrate that we are able to increase the effective quantum volume from three to five on Belem, three to four on Lima, and from four to five on Quito. 
Note that ZNE increases the estimated heavy output probability $\hat{h}_d$ on all qubit subsets even though the $2/3$ threshold is not always crossed. We therefore expect that ZNE can increase effective quantum volume independent of the size of the device, so long as cross-talk and other errors do not scale with the device size. The largest ZNE experiment to date was performed on $26$ qubit circuits with $1080$ two-qubit gates~\cite{zne_experiment_2021} and, in the context of our work, provides evidence that error mitigation can continue to increase the effective volume of larger quantum devices, e.g. those listed in Appendix~\ref{app:measured-qvs}.

Because we estimate the noiseless result by taking a linear combination of noisy results, the way we compute $\sigma$ in~\eqref{eqn:volume-condition} changes relative to~\cite{qv}. For any technique, such as Richardson extrapolation, that evaluates an error-mitigated expectation value as a linear combination of noisy expectation values~\eqref{eqn:zne-result}, one can show (see Appendix~\ref{app:statistical-uncertainty-of-error-mitigated-volume}) that
\begin{align}
    \sigma^2 = \frac{1}{n_c^2}\sum_{C} \sigma_C^2, \qquad  
    & \sigma_C^2 =  \sum_{i = 1}^{k} |\eta_i|^2 ( \sigma^{(i)}_C)^2 ,
\end{align}
where $(\sigma^{(i)}_C)^2$ is the variance of each noise-scaled expectation value, while $\sigma_C^2$ is the variance of
the error-mitigated expectation value associated to the quantum circuit $C$.
The previous expressions correspond to a theoretical estimate of the error, but in practice we can estimate error bars by repeating the experiment multiple times or by bootstrapping. The $2\sigma$ error bars in Fig.~\ref{fig:main} are calculated by bootstrapping with $500$ resamples. See Appendix~\ref{app:statistical-uncertainty-of-error-mitigated-volume} for more details. % and other ways to compute error bars.

\textbf{Discussion} \ \ 
There is a subtle point in interpreting our results in the general context of quantum computer performance. 
Our error mitigation procedure improves the expectation value of the heavy output projector but does not produce more heavy bitstrings --- in fact, our procedure likely produces fewer heavy bitstrings because we distribute samples across circuits at amplified noise levels. However, as we have shown, we are able to use this information to estimate the expected number of heavy bitstrings in a statistically significant way. 
To carefully distinguish between the two cases, we refer to our results as increasing the \textit{effective} quantum volume.

%Evaluating~\eqref{eqn:zne-result} allows us to estimate the mean value and standard deviation of a random variable corresponding to {how many} heavy bitstrings were sampled, but it does not tell us {which} heavy bitstrings were sampled. While the quantum volume conditions~\eqref{eqn:volume-condition} and many quantum algorithms do not require us to know which bitstrings were sampled, some algorithms do require this information. To carefully distinguish between the two cases, we refer to our results as increasing the \textit{effective} quantum volume, even though they pass the quantum volume test as it is defined in~\cite{qv}. 

The restriction to evaluating expectation values but not directly sampling bitstrings raises interesting questions about physicality and the role of a quantum computer in a computational procedure.  If an algorithm only requires expectation values and we apply the error mitigation procedure used in this work, is it the case that we effectively have access to a quantum computer with a larger quantum volume? These questions are linked to the physical interpretation of error mitigation. One way to interpret ZNE is that we evaluate expectation values with respect to the ``extrapolated density matrix''
\begin{equation}
    \rho_0 = \sum_i \eta_i \rho_{\lambda_i}
\end{equation}
where $\rho_{\lambda_i}$ are the noise-scaled physical states and $\eta_i$ are the real coefficients in~\eqref{eqn:zne-result}. Clearly we did not physically prepare $\rho_0$ in our experiment, but should we restrict the use of a quantum computer to only preparing a single physical state from which we can sample bitstrings? Or do we allow ourselves to ``virtually'' prepare non-physical but mathematically well-defined states from which we can compute expectation values more accurately? We note that similar questions have been asked in the context of virtual distillation techniques~\cite{vd, koczor} which have been proposed to artificially purify a quantum state or to reduce its effective temperature~\cite{qvc}.

\textbf{Conclusion} \ \ 
In this work we have experimentally demonstrated that error mitigation improves the effective quantum volume of several quantum computers. We use the term \textit{effective} quantum volume to emphasize that our procedure is appropriate for algorithms computing expectation values and not for algorithms requiring individual bitstrings.
The error mitigation technique is not tailored to the structure of quantum volume circuits or to the architecture of the quantum computers we used. Indeed, we did not run any additional calibration experiments or use any calibration information to obtain our results. Similar software-level techniques have been used in previous quantum volume experiments, e.g. (approximate) compilation~\cite{qv, qv_compile} and dynamical decoupling~\cite{qv_compile}. The novelty of our proposal is that, by relaxing the strong requirement of directly sampling heavy bitstrings to the weaker requirement of estimating the expectation value of the heavy output projector, more general error mitigation techniques can be applied to improve the effective quantum volume of a device. We expect this approach to improve the effective quantum volume of additional quantum computers such as those in Appendix~\ref{app:measured-qvs}. Our open source error mitigation software~\cite{mitiq} can be used on many quantum computers to repeat the experiments we performed here.

% In this work we have experimentally demonstrated that error mitigation improves the effective quantum volume of several quantum computers. We use the term \textit{effective} quantum volume to emphasize that our procedure is appropriate for algorithms computing expectation values and not for algorithms requiring individual bitstrings. The error mitigation technique is not tailored to the structure of quantum volume circuits or to the architecture of the quantum computers we used. Indeed, we did not run any additional calibration experiments or use any calibration information to obtain our results. While error mitigation does use additional resources (here, additional gates in some of the compiled circuits), this is permissible in the definition of quantum volume~\cite{qv} and an analogous technique has been used in previous quantum volume experiments: namely, additional gates were added to circuits for dynamical decoupling in~\cite{qv_compile}. We expect error mitigation as used in our work to improve the effective quantum volume of additional quantum computers such as those in Appendix~\ref{app:measured-qvs}. Our open source error mitigation software~\cite{mitiq} can be used on many quantum computers to repeat the experiments we performed here.

In the context of error mitigation, our work provides additional benchmarks to the relatively few experimental results in literature~\cite{zne_experiment_2019, mitiq, zne_experiment_2021, pec_experiment_2020, pec_experiment_2022, lanl_em_benchmarks, huffman}. We encourage the use of quantum volume as a benchmark for error mitigation techniques due to its relatively widespread adoption and clear operational meaning. Normalizing by additional resources used (gates, shots, qubits, etc.) in error-mitigated quantum volume experiments provides a way to directly compare different techniques and drive progress in this area. As most error mitigation techniques act on expectation values, they can be used for effective quantum volume experiments as we have done in this work. 

\vspace{0.5em}

\textbf{Code and data availability} The code we used to run experiments as well as the data we collected are available at \href{https://github.com/unitaryfund/mitiq-qv}{https://github.com/unitaryfund/mitiq-qv}.

\vspace{0.5em}

\textbf{Acknowledgements} This work was supported by the U.S. Department of Energy, Office of Science, Office of Advanced Scientific Computing Research, Accelerated Research in Quantum Computing under Award Number DE-SC0020266. R.L. acknowledges support from a NASA Space Technology Graduate Research Fellowship. We thank IBM for providing access to their quantum computers and software for reproducing quantum volume experiments. The views expressed in this paper are those of the authors and do not reflect those of IBM.

% We have shown by means of other benchmarks that zero-noise extrapolation is not specifically tailored to quantum volume circuits or the particular quantum computer architecture used in our experiments. We anticipate that other general-purpose error mitigation techniques can be used in a similar fashion to improve quantum volume. While there are many benchmarks that can be used to assess the performance of an error mitigation technique, we encourage the adoption of the quantum volume benchmark for this purpose as it is a relatively standard measure that offers direct comparison between different techniques. 

% While ZNE does use additional resources (i.e., additional gates in noise-scaled circuits), this can be considered a type of compiling which is permissible in quantum volume experiments...

% Zero-noise extrapolation is not tailored to the particular structure of quantum volume circuits in any way and can be used on any circuit with invertible gates and terminal measurements. To demonstrate this we perform ZNE on various other benchmark circuits and observables. The results, shown in Fig.~[], indicate that ZNE is reliable for other types of random benchmark circuits as well as structured circuits. 

% Additionally, ZNE is not tailored to any particular quantum computer architecture or qubit type... 

\bibliography{refs.bib}

% \newpage

\appendix

\section{Device specifications} \label{app:device-specifications}

In Table~\ref{tab:devices} we provide more information about the quantum computers we used in our experiments. Note that the quantum volume of Belem is listed as four at~\cite{ibmq} but we are unable to reproduce this result in our unmitigated experiments, presumably due to device degradation over time.

\begin{table}[h!]
    \centering
    \begin{tabular}{c|c|c|c}
                                 & Lima                   & Belem                  & Quito                  \\ \hline 
        \# Qubits                & 5                      & 5                      & 5                      \\
        $\epsilon_{\text{1Q}}$   & $4.446 \times 10^{-4}$ & $2.808 \times 10^{-4}$ & $2.980 \times 10^{-4}$ \\
        $\epsilon_{\text{CNOT}}$ & $1.131 \times 10^{-2}$ & $1.098 \times 10^{-2}$ & $8.292 \times 10^{-3}$ \\
        $\epsilon_{\text{M}}$    & $3.790 \times 10^{-2}$ & $2.868 \times 10^{-2}$ & $2.546 \times 10^{-2}$ \\
    \end{tabular}
    \caption{Device specifications and error rates for the quantum computers we used in our experiments. Device connectivities are shown in Fig.~\ref{fig:main}. Parameters $\epsilon_{\text{1Q}}$, $\epsilon_{\text{CX}}$, $\epsilon_{\text{M}}$ denote, respectively, averages (over all qubits) of single-qubit $\sqrt{X}$ gate errors, two-qubit CNOT gate errors, and readout errors $(p(0|1) + p(1|0)) / 2$ accessed from~\cite{ibmq}.}
    \label{tab:devices}
\end{table}

% \newpage % This is just to get the table in the correct section. Can be removed.
\section{Table of quantum volumes} \label{app:measured-qvs}

As discussed in the main text, error mitigation consistently increased the estimated heavy output probability in all of our experiments. To increment the effective quantum volume of a device, this increase must cross the $2/3$ threshold with statistical significance. While there is no guarantee that this will happen, we expect there to be several cases of already-measured quantum volumes where this will be true. For this reason, as well as general context, we include a list of quantum computer volumes in Table~\ref{tab:qvs}. 

\begin{table}[h!]
    \centering
    \begin{tabular}{c|c|c}
        Quantum computer & $\log \text{QV}$  & Reference \\ \hline
        Rigetti Aspen-4 & 3 & \cite{rigetti_qv3} \\
        \textcolor{black}{Lima} & 3 (4) & \cite{ibmq} (this work) \\
        \textcolor{black}{Belem} & 3 (5) & \cite{ibmq} (this work) \\
        \textcolor{black}{Jakarta} & 4 & \cite{ibmq} \\
        Bogota & 4 & \cite{ibmq} \\
        \textcolor{black}{Quito} & 4 (5) & \cite{ibmq} (this work) \\
        Manila & 5 & \cite{ibmq} \\
        Nairobi & 5 & \cite{ibmq} \\
        Lagos & 5 & \cite{ibmq} \\
        Perth & 5 & \cite{ibmq} \\
        Guadalupe & 5 & \cite{ibmq} \\
        Toronto & 5 & \cite{ibmq} \\
        Brooklyn & 5 & \cite{ibmq} \\
        Trapped-ion QCCD & 6 & \cite{qccd_qv6} \\
        Hanoi & 6 & \cite{ibmq} \\
        Auckland & 6 & \cite{ibmq} \\
        Cairo & 6 & \cite{ibmq} \\
        Washington & 6 & \cite{ibmq} \\
        Mumbai & 7 & \cite{ibmq} \\
        Kolkata & 7 & \cite{ibmq} \\
        Honeywell System Model H1 & 10 & \cite{honeywell_qv10} \\
        % IonQ 5th Generation System & 22$^*$ & \href{https://www.nextbigfuture.com/2021/03/ionq-quantum-computer-4-million-quantum-volume-and-16x-error-correction.html}{url} \\
    \end{tabular}
    \caption{Measured quantum volumes (in increasing order). Values in parentheses show effective quantum volumes measured in this work.
    %We increased the effective quantum volume of \textcolor{red}{red entries} with error mitigation in this work.
    % $^*$Estimated, not measured.
    }
    \label{tab:qvs}
\end{table}

\section{\label{app:statistical-uncertainty-of-error-mitigated-volume} Statistical uncertainty of error-mitigated volume}

%\begin{figure}
%    \centering
%    \includegraphics[width=\columnwidth]{quito_final_empirical_std.pdf}
%    \caption{Quito quantum volume experiment (middle panel of Fig.~\ref{fig:main}) with $\sigma$ estimated by dividing the $n_c = 500$ circuits into five groups of $100$ circuits. As in the main text, error bars show $2\sigma$.}
%    \label{fig:my_label}
%\end{figure}

\subsection{Theoretical estimation of error bars}

For a large number of error mitigation techniques, including Richardson extrapolation,
an error-mitigated expectation value $E_C$ associated with an ideal circuit $C$ is evaluated as linear combination of different noisy expectation values:
\begin{equation}
    E_C = \sum_{j} \eta_j \tilde{E}_j.
\end{equation}
Because of shot noise, each noisy expectation value $\tilde E_j$ can only be measured up to a statistical variance $\sigma_j^2 =  \mathbb{E}( \tilde E_j^2) - [\mathbb{E}(\tilde E_j)]^2$, where $\mathbb{E}$ represents the statistical average over $n_j$ measurement shots.

Since different noisy expectation values are statistically uncorrelated,
the variance $\sigma_C^2$ of the error-mitigated result $E_C$ is:

\begin{equation}
    \sigma_C^2 =  \mathbb{E}(E_C^2) - [\mathbb{E}( E_C)]^2 = \sum_j |\eta_j|^2 \sigma_j^2.
\end{equation}

If we assume that each noisy expectation value is obtained by sampling a binomial distribution $\mathcal B(p_j, n_j)$ with probability $p_j=\tilde E_j$ and normalizing the result over $n_j$ measurement shots, we have $\sigma_j^2 = E_j (1 - E_j)/n_j $. The variance $\sigma_C^2$ of the error-mitigated result is therefore:
\begin{equation} \label{eq:app_sigma_c}
    \sigma_C^2 =  \sum_{j=1}^k |\eta_j|^2 \tilde E_j (1 - \tilde E_j)/n_j.
\end{equation}
The previous expression is valid for a generic  expectation value. In the specific case of a quantum volume experiment, we can identify with $E_C$ the heavy-output probability associated with a specific random circuit $C$. 
Averaging $E_C$ over multiple $n_c$ noisy circuits $C$ of depth $d$, produces the estimated heavy output probability visualized in Fig. \ref{fig:main}:
\begin{equation} \label{eq:app_hd}
h_d = \frac{1}{n_c}\sum E_C.
\end{equation}
This is again a sum of independent random variables and so its variance is given by:

\begin{equation} \label{eq:app_sigma}
    \sigma^2 = \frac{1}{n_c^2}  \sum_{C}  \sigma_C^2.
\end{equation}

%In a quantum volume experiment, a conservative claim of success can be defined by the worst-case condition
%$E - z \sigma > 2/3$, where $z$ is the number of standard deviations that we want to consider in the estimation (e.g. $z=2$). 
%In particular, in a quantum volume experiment, $E_j$ is estimated as $n_h^{(j)} / (n_c n_s)$ with where $n_h^{(j)}$ is the number 
%of heavy bitstrings $n_c$ is the number of circuits and $n_s$ is the number of shots per circuit. In this case we have:

%\begin{equation}
%    \sigma =  \frac{1}{n_c n_s} \sqrt{\sum_j |\eta_j|^2 n_h^{(j)} (1 - \frac{n_h}{n_c n_s} )}.
%\end{equation}

%A more conservative estimate of the standard deviation, which takes into account potential systematic errors, is the following~\cite{qv}:

%\begin{equation}
%    \sigma' = \sqrt{n_s} \sigma =  \frac{1}{n_c n_s} \sqrt{\sum_j |\eta_j|^2 n_h (n_s - \frac{n_h}{n_c} )} \ge \sigma.
%\end{equation}

\subsection{Bootstrapping empirical error bars}

\begin{figure}
    \centering
    \includegraphics[width=\columnwidth]{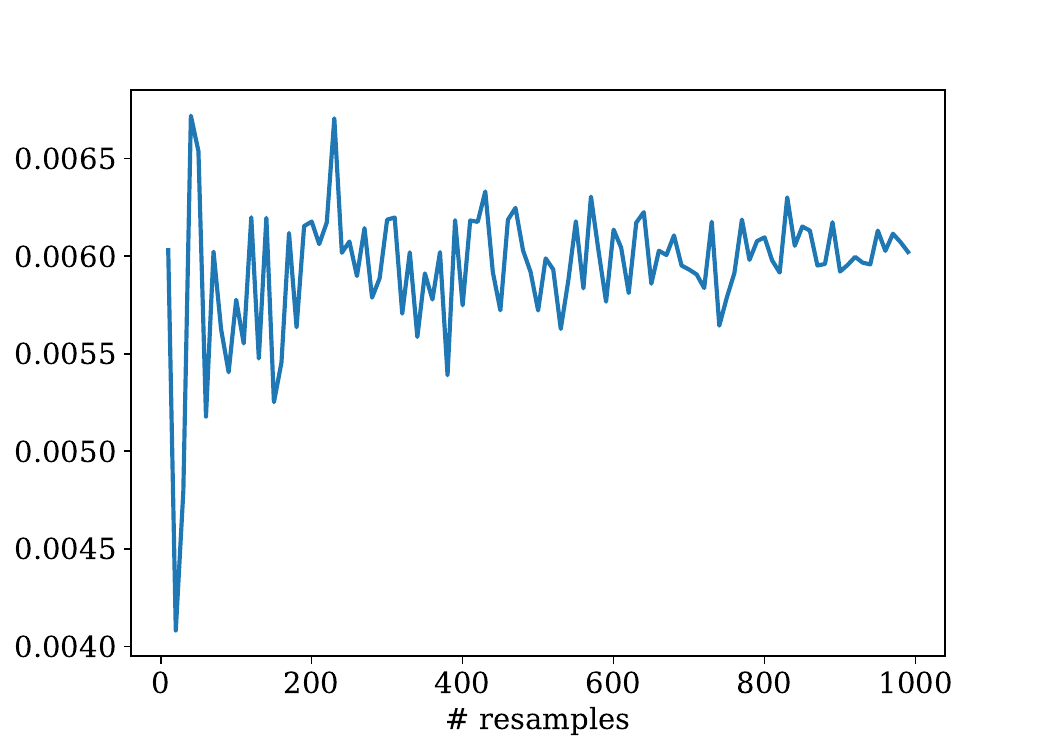}
    \caption{The value of $\sigma$ for different resampling numbers in bootstrapping.}
    \label{fig:bootstrap}
\end{figure}

The previous way of estimating error bars is based on theoretical assumptions and, even though it provides useful analytical expressions, it may underestimate unknown sources of errors such as systematic errors.

A brute-force way of estimating error bars is to repeat the estimation of the quantity of interest (in our case $h_d$) with $N$ independent  experiments and to evaluate the empirical variance of the results. This method can be expensive with respect to classical and quantum computational resources, and the results can be sensitive to how the independent samples are grouped. So while we can split independent samples into five groups of $n_c = 100$ circuits to estimate the standard deviation this way, a more feasible alternative is instead given by {bootstrapping}. This is  a statistical inference technique in which, instead of performing $N$ new  experiments, one  {\it resamples} the raw results of a single experiment $N$ times in order to estimate properties of the underlying statistical distribution.

In our specific quantum volume experiment, the heavy-output probability $h_d$ is estimated as an average over $n_C$ random circuits $C$ as shown in equation \eqref{eq:app_hd}. Let us define the set $S = \{ E_{C_1}, E_{C_2}, .... E_{C_{n_C}} \}$ containing all the estimated heavy-output probabilities associated with different random circuits.
We can now resample N sets of data $S_1, S_2, .... S_N$, each one containing $n_c$ values that are randomly sampled from $S$ with replacements. For each resampled set $S_j$ we  evaluate the associated  bootstrapped mean 
$\mu_j =  \langle E_C \rangle_{S_j}$.

The empirical standard deviation of all the bootstrap means  $\{\mu_1, \mu_2, ... \mu_N\}$ provides an estimate of the statistical uncertainty:

\begin{equation}
\sigma =\sqrt{ \frac{1}{N} \sum_{j=1}^N (\mu_j - \bar \mu)^2}.
\end{equation}
where $\bar \mu = \sum_{j=1}^N \mu_j/N$.
This is the method that we used to evaluate error bars in Fig. \ref{fig:main} (with  $N=500$).

One may ask how large should $N$ be, in order to obtain a resonable estimate of the error. In Fig. \ref{fig:bootstrap} we show the dependance of $\sigma$ for an arbitrary error-mitigated point of Fig. \ref{fig:main} (the results are qualitatively similar for all points). Fig. \ref{fig:bootstrap} provides empirical evidence that, for $N>400$, the bootstrapped estimate converges around a stable result.
\end{document}